\documentclass[aps,prc,12pt,showpacs,showkeys,onecolumn,superscriptaddress,groupedaddress]{revtex4-1}

\usepackage[english]{babel}
\usepackage{indentfirst}
\usepackage{amssymb}
\usepackage{braket}
\usepackage{bm}
\usepackage{amsxtra}
\usepackage{amsmath}
\usepackage{supertabular}
\usepackage{multirow}
\usepackage{color}
\usepackage [mathcal]{eucal}
\usepackage{graphicx}
\usepackage{graphics}
\usepackage{lipsum}
\usepackage{comment}

\def\n4lo{$\mathrm{N}^4\mathrm{LO}$}
\def\n3lo{$\mathrm{N}^3\mathrm{LO}$}
\def\dblone{\hbox{$1\hskip -1.2pt\vrule depth 0pt height 1.6ex width 0.7pt
     \vrule depth 0pt height 0.3pt width 0.12em$}}
     \def\chiral4lo{$\mathrm{N}^4\mathrm{LO}$}

\begin{document}

\noindent
\title{Impact of Three-Body Forces on Elastic Nucleon-Nucleus Scattering Observables}

\author{Matteo Vorabbi$^{1}$}
\author{Michael Gennari$^{2,3}$}
\author{Paolo Finelli$^{4}$}
\author{Carlotta Giusti$^{5}$}
\author{Peter Navr\'atil$^{3}$}
\author{Ruprecht Machleidt$^{6}$}

\affiliation{$~^{1}$National Nuclear Data Center, Bldg. 817, Brookhaven National Laboratory, Upton, NY 11973-5000, USA
}

\affiliation{$~^{2}$University of Victoria, 3800 Finnerty Road, Victoria, British Columbia V8P 5C2, Canada
}

\affiliation{$~^{3}$TRIUMF, 4004 Wesbrook Mall, Vancouver, British Columbia, V6T 2A3, Canada
}

\affiliation{$~^{4}$Dipartimento di Fisica e Astronomia, 
Universit\`{a} degli Studi di Bologna and \\
INFN, Sezione di Bologna, Via Irnerio 46, I-40126 Bologna, Italy
}

\affiliation{$~^{5}$Dipartimento di Fisica,  
Universit\`a degli Studi di Pavia and \\
INFN, Sezione di Pavia,  Via A. Bassi 6, I-27100 Pavia, Italy
}

\affiliation{$~^{6}$Department of Physics, University of Idaho, Moscow, Idaho 83844, USA
}

\date{\today}


\begin{abstract} 

{\bf Background:} In a previous series of papers we investigated the domain of applicability of chiral potentials to the construction of a microscopic optical potential (OP) for
elastic nucleon-nucleus scattering. The OP was derived at the first order of the spectator expansion of the Watson multiple scattering theory and its final expression was a folding
integral between the nucleon-nucleon ($NN$) $t$ matrix and the nuclear density of the target.  In the calculations $NN$ and three-nucleon ($3N$) chiral interactions were used for
the target density and only the $NN$ interaction for the $NN$ $t$ matrix.

{\bf Purpose:} The purpose of this work is to achieve another step towards the calculation of a more consistent OP introducing
the $3N$ force also in the dynamic part of the OP.

{\bf Methods:} The full treatment of the $3N$ interaction is beyond our present capabilities.
Thus, in the present work it is approximated with a density dependent $NN$ interaction obtained after the averaging over the Fermi sphere. In practice, in our
model the $3N$ force acts as a medium correction of the bare $NN$ interaction used to calculate the $t$ matrix. Even if the $3N$ force is treated in an approximate way, this method
naturally extends our previous model of the OP and allows a direct comparison of our present and previous results. 

{\bf Results:} We consider as case studies the elastic scattering of nucleons off $^{12}$C and $^{16}$O. We present results for the differential cross section and the spin observables for different values of the projectile energy. From the comparison with the experimental data and with the results of our previous model  
we assess the
importance of the $3N$ interaction in the dynamic part of the OP.

{\bf Conclusions:} Our analysis indicates that the contribution of the $3N$ force in the $t$ matrix is small for the differential cross section and it is  sizable for  the spin observables, in particular, 
for the analyzing power. 
We find that the two-pion exchange term is the major
contributor to the $3N$ force. A chiral expansion order-by-order analysis of the scattering observables 
confirms the convergence of our results at the next-to-next-to-next-to-leading-order, as already established in our previous work.

\end{abstract}

\pacs{24.10.Ht;25.40.Cm;25.40.Dn;24.70.+s;11.10.Ef}

\maketitle

\section{Introduction}

The optical potential (OP) is a widely used tool developed in the first instance to describe the elastic nucleon-nucleus scattering and successively employed in other
nuclear reactions. Decades of research work have led to the development of different phenomenological and microscopic approaches to derive OPs  to be employed in different
kinematical regions and for different reactants.
A phenomenological approach is generally preferred to achieve a more accurate description of the available experimental data. 
Despite this accuracy, the predictive power of phenomenological OPs remains poor when they are applied to situations for which data are not yet available, due to their dependence
on several free parameters fitted to reproduce the existing data. A microscopic approach to the OP still remains the preferred way to make reliable predictions and to assess the
impact of the approximations introduced in the model, and, recently, several new works have been devoted to this topic~\cite{dickhoff_2017,DICKHOFF2019,PhysRevC.101.044303,Idini2017,PhysRevLett.123.092501,PhysRevC.95.024315,PhysRevC.98.044625,rotureau_fphy.2020.00285,PhysRevC.100.014601,PhysRevC.101.064613,Whitehead:2020wwb,durant_2018,Arellano_2018,arellano_2019,Arellano_2020,PhysRevC.92.024618,*PhysRevC.96.059905,Kohno_ptep_pty001,PhysRevC.102.024611,PhysRevC.99.034605,PhysRevC.100.014613,Burrows:2018ggt,Burrows:2020qvu,Gennari:2017yez,Vorabbi_2020}.

At intermediate energies, the construction of a microscopic OP based on the Watson multiple scattering theory is particularly appealing, and in the 90's it produced several
theoretical works~\cite{PhysRevC.41.2257,PhysRevC.46.279,PhysRevC.50.2995,PhysRevC.41.2188,*PhysRevC.42.1782,PhysRevC.42.652,PhysRevC.52.301,PhysRevC.41.814,PhysRevC.56.2080} where realistic nucleon-nucleon ($NN$) interactions together with nuclear target densities were used as the input for the calculation of such
microscopic OPs.

The development of new $NN$ and three-nucleon ($3N$) interactions derived within the framework of the chiral perturbation theory (ChPT), together with the modern
accurate many-body techniques, resulted in a renewed interest in the subject, because of the possibility to achieve a more consistent calculation of the OP using the $NN$ and $3N$
forces as the only input for the computation of its dynamic and structure parts. We note that this choice is not unique and, recently, a similar OP has been successfully
derived~\cite{Burrows:2018ggt,Burrows:2020qvu} using only the $NN$ interaction, in particular the one from Ref.~\cite{PhysRevLett.110.192502}.

In a series of papers we explored the possibility of constructing a microscopic OP from chiral interactions: we derived a microscopic OP from $NN$ chiral
potentials~\cite{Vorabbi:2015nra}, we studied the convergence of the scattering
observables computed with $NN$ potentials at different chiral orders~\cite{Vorabbi:2017rvk}, we investigated the predictive power of our OP against the experimental
data for several isotopic chains~\cite{Vorabbi_2018} and compared our results with those obtained with one of the most popular phenomenological
OP~\cite{Koning:2003zz,KONING2014187}. Our original model was improved in Ref.~\cite{Gennari:2017yez}, where we computed our OP with a microscopic nonlocal density
obtained with the {\it ab initio}  no-core shell model (NCSM)~\cite{BARRETT2013131} utilizing $NN$ and $3N$ chiral interactions. The same $NN$ interaction was used in
Ref.~\cite{Gennari:2017yez} to calculate the $NN$ $t$ matrix and the nuclear density, that convoluted together give the OP. Recently, this approach has been also extended to
describe the elastic scattering of antiprotons off several target nuclei~\cite{Vorabbi_2020}.

Despite all these advances, a lot of work has still to be done before reaching  full consistency. In particular, the approach adopted in
Ref.~\cite{Gennari:2017yez} uses $NN$ and $3N$ interactions to calculate the nuclear density, while the $NN$ $t$ matrix, which represents the dynamic
part of the OP, is computed with the $NN$ interaction only.
Naively, we can argue that the impact of the $3N$ force is more important in the nuclear density, since reproducing the
nuclear radii is essential for a proper description of the diffraction minima in the differential cross section. However, for a more consistent derivation, the $NN$ and $3N$ potentials
should be used both in the dynamic and in the structure parts of the OP. Unfortunately, the exact treatment of the $3N$ interaction 
is a very hard task that is beyond our present capabilities.

The goal of the present work is to develop a framework that allows us to introduce and consequently assess the impact of the $3N$ force in the dynamic
part of the OP. Our framework makes use of a density-dependent $NN$ interaction, which introduces some medium corrections to the bare $NN$ potential in the calculation of
the $t$ matrix and naturally extends the previous scheme adopted in Ref.~\cite{Gennari:2017yez}.

The paper is organized as follows: in Sec.~\ref{sec_optical_model} we derive the expression for the OP operator and we show explicitly how the $3N$ force is included in our scheme. Some details about the chiral potentials can be found in Sec.~\ref{Sec3B}, while the technical details about the calculation of the OP are given in Sec.~\ref{sect_op}.
In Sec.~\ref{sec_theoretical_predictions} we show the results for the scattering observables obtained with our OP and compare them to the experimental data. Finally,
in Sec.~\ref{sec_conclusions} we draw our conclusions.

\section{Optical Model}
\label{sec_optical_model}
 
In the most general framework $3N$ effects could arise both at the level of the bare nuclear potential or as a result of the complicated many-body dynamics. 
Recalling the distinction introduced by Sauer in Ref. ~\cite{Sauer_2014}, many-nucleon forces can be generally divided into two categories: {\em genuine} contributions arising
from the nuclear Hamiltonian and {\em induced} terms coming from the process of solving the nuclear many-body problem. 
Induced many-nucleon forces do not have a fundamental basis. In some sense they can be interpreted as theoretical artifacts due to the inevitable approximations involved in the solution of the many-body problem. On the other hand, genuine contributions enter directly into the definition of the nuclear Hamiltonian
in terms of the active degrees of freedom chosen to describe the nuclear systems.

Our aim is to present a consistent framework in which the role of $3N$ forces in elastic nucleon-nucleus scattering can be investigated. In this perspective, we will restrict our
analysis to the role of genuine $3N$ forces and neglect, as a first step, the complications due to induced many-body forces. 
Since at the moment an exact treatment of the full problem is not available, we will focus our attention on the impact of $3N$ forces in the approximation
of $NN$ dynamics dominance.

We will mainly follow the derivation presented in 
Refs.~\cite{Vorabbi:2015nra,Vorabbi:2017rvk, 1997PhDT67W, PhysRevC.48.2956, PhysRevC.51.1418, PhysRevC.52.1992,PhysRevC.56.2080} in order to assess the
strengths and limitations of our analysis.

To deal with the general problem of the elastic scattering of a nucleon from a target nucleus of $A$ nucleons, we start from the full $(A+1)$-body Lippmann-Schwinger equation for
the many-body transition amplitude $T$ as follows
\begin{equation}
T = V + V G_0 (E) T \; ,
\label{LSeq}
\end{equation}
where $V$ is the chiral nuclear potential at a given order in the relevant expansion parameter (more details in Sec.~\ref{Sec3B}) and $G_0(E)$ is the $(A+1)$-body propagator
connected to the free Hamiltonian $H_0$ (that includes the target Hamiltonian $H_A$ and the kinetic energy of the projectile $h_0$) defined as
\begin{equation}
G_0(E) = (E - h_0 -H_A + i \epsilon)^{-1} \; .
\end{equation}
In the standard approach to elastic scattering, Eq.~(\ref{LSeq}) is separated into two equations. The first one is an integral equation for $T$
\begin{equation}\label{firsttamp}
T = U + U G_0 (E) P T \, ,
\end{equation}
where $U$ is the optical potential operator, and the second one is an integral equation for $U$
\begin{equation}\label{optpoteq}
U = V + V G_0 (E) Q U \, .
\end{equation}
In the previous expressions we introduced the projection operators $P$ and $Q$ that satisfy the relation
\begin{equation}
P + Q = \dblone \, ,
\end{equation}
and where $P$ fulfills the condition
\begin{equation}\label{procommutator}
[G_0 , P] = 0 \, .
\end{equation}
Of course, in the case of elastic scattering, $P$ projects onto the elastic channel. It can be defined as follows
\begin{equation}
P = \frac{\ket{\Phi_A} \bra{\Phi_A}}{\braket{\Phi_A|\Phi_A}} \, ,
\end{equation}
where $\ket{\Phi_A}$ is the ground state of the target.
With these definitions, the elastic transition operator may be defined as $T_{\mathrm{el}} = PTP$, and, in this
case, Eq.~(\ref{firsttamp}) becomes
\begin{equation}\label{elastictransition}
T_{\mathrm{el}} = P U P + P U P G_0 (E) T_{\mathrm{el}} \, .
\end{equation}
Thus the transition operator for elastic scattering is given by a one-body integral equation. In order to solve
Eq.~(\ref{elastictransition}) we need to know the operator $P U P$.

The nuclear potential derived in the framework of ChPT consists of two-body and, starting at N$^{2}$LO (next-to-next-to-leading order) in the perturbative expansion, also three-body
contributions, see Refs. ~\cite{BERNARD_1995, VanKolck1999337, chiralmachleidt_n3lo, chiralepelbaum_n3lo, Epelbaum:2008ga, Machleidt:2011zz, Epelbaum:2014sza, Epelbaum:2014efa, Entem:2014msa, Entem:2017gor}. As well known, the $NN$ potentials at order LO (leading order) and NLO (next-to-leading order) are not a viable choice since,
as we have shown in our previous papers ~\cite{Vorabbi:2015nra,Vorabbi:2017rvk}, the scattering observables are poorly reproduced. 
Starting from this consideration, it is useful to study the effects of the $3N$ force in the solutions of Eq. (\ref{LSeq}) because, in addition to the aforementioned nuclear potential $V$,
the inclusion of the $3N$ force is an essential piece in the {\it ab initio} description of nuclear targets (with the exception of the potential
NNLO$_{\mathrm{sat}}$, see  Ref.~\cite{PhysRevC.91.051301}, that, however, is not suited to be employed in proton elastic scattering at energies larger
than 100 MeV ~\cite{machleidt2019wrong}). 

Let us start by writing the chiral potential as follows,
\begin{equation}
V  = V_{NN} + V_{3N} \; ,
\label{V_with_3B}
\end{equation}
where $V_{NN}$ consists of all two-body contributions $v_{0i}$ between the nucleon projectile (labelled by $0$) and the {\em i}th nucleon in the target,
\begin{equation}
V_{NN} = \sum_{i=1}^A v_{0i} \; ;
\end{equation} 
and $V_{3N}$ is determined by all three-body contributions $w_{0ij}$ between the projectile and two spectator nucleons in the target ($i$ and $j$),
\begin{equation}
V_{3N} = \frac{1}{2}\sum_{i=1}^A \sum_{j=1\; j \ne i}^A w_{0ij} \; .
\end{equation}
Now we insert Eq.~(\ref{V_with_3B}) into Eq.~(\ref{optpoteq}) and we obtain the many-body equation
for the optical potential operator
\begin{equation}\label{total_op_equation}
U = ( V_{NN} + V_{3N} ) + ( V_{NN} + V_{3N} ) G_0 (E) Q U \, .
\end{equation}
The exact treatment and solution of the previous equation is beyond our current capabilities already with the $NN$ interaction only, so, in order to include some effects due to
a $3N$ force, we need to introduce an approximation which allows us to simplify the previous equation to a form that can be treated with the standard techniques.
If we make the assumption that the two-nucleon dynamics dominates the scattering processes, we can introduce the following approximation
\begin{equation}\label{v_3Nf}
\sum_{\substack{j=1 \\j \neq i}}^A w_{0ij} \approx \braket{w_{0i}} \, ,
\end{equation}
where the notation $\braket{\ldots}$ indicates an average over the Fermi sphere. The operator  $\braket{w_{0i}}$ is a two-body operator.
How to perform such a simplification will be described in Sec.~\ref{Sec3B}.
If we insert Eq.~(\ref{v_3Nf}) into Eq.~(\ref{total_op_equation})
and we define the following potentials
\begin{align}
v_{0i}^{(1)} &\equiv v_{0i} + \frac{1}{2} \braket{w_{0i}} \, , \\
v_{0i}^{(2)} &\equiv v_{0i} + \braket{w_{0i}} \, ,
\end{align}
we obtain
\begin{equation}\label{op_one_body}
U = \sum_{i=1}^A U_{0i} \, ,
\end{equation}
where
\begin{equation}\label{LS_one_body_op_eq}
U_{0i} = v_{0i}^{(1)} + v_{0i}^{(2)} G_0 (E) Q U \, .
\end{equation}
If we insert Eq.~(\ref{op_one_body}) into Eq.~(\ref{LS_one_body_op_eq}) and we define the following operators
\begin{align}
\tau_{0i} &\equiv v_{0i}^{(1)} + v_{0i}^{(2)}  G_0 (E) Q \tau_{0i} \, , \\
\chi_{0i} &\equiv v_{0i}^{(2)} + v_{0i}^{(2)}  G_0 (E) Q \chi_{0i} \, ,
\end{align}
we obtain
\begin{equation}\label{second_one_body_op_eq}
U_{0i} = \tau_{0i} + \chi_{0i} G_0 (E) Q \sum_{\substack{j=1 \\j \neq i}}^A U_{0j} \, .
\end{equation}
We see that the operator $\tau_{0i}$ satisfies a Lippmann-Schwinger equation and is density dependent, because of the presence of the operator $\braket{w_{0i}}$.
In the limit of the density going to zero the operator $\tau_{0i}$ becomes equal to the first-order term of the spectator expansion. 
Our approach explicitly neglects contributions from higher-order terms in the spectator expansion that would naturally produce induced three-body forces.
We can now approximate Eq.~(\ref{second_one_body_op_eq}) with its leading term and, since this is still a many-body equation, we introduce the impulse approximation.
Thus, after some manipulations ~\cite{Vorabbi:2015nra,Vorabbi:2017rvk}, the final expression for the optical potential is given by
\begin{equation}
U = \sum_{i=1}^A t_{0i} \, ,
\end{equation}
where
\begin{equation}
\label{final_t}
t_{0i} = v_{0i}^{(1)} + v_{0i}^{(2)}  g_i t_{0i}
\end{equation}
and
\begin{equation}
g_i = \frac{1}{(E - E_i ) - h_0 - h_i + i \epsilon} \, .
\end{equation}
Here we see that $g_i$ is the two-body free propagator while, in the limit of a zero density, the operator $t_{0i}$ becomes the free two-body scattering operator.

Essentially, what we did is to approximate the pure $3N$ force with a density-dependent $NN$ force obtained by averaging the third nucleon momenta over the Fermi sphere.
Within this procedure, we produce a term in Eq.~(\ref{final_t}) that introduces {\it de facto} a medium correction to the standard expression of the OP obtained in the
impulse approximation. Our treatment of medium corrections is not exhaustive and other contributions can also be included, as suggested in
Refs. ~\cite{PhysRevC.52.1992, PhysRevC.88.064005, PhysRevC.92.024618, PhysRevC.98.054617, PhysRevC.52.301}.
A more complete investigation of medium corrections is mandatory for the future of our model and we plan to investigate a complete treatment  in a forthcoming article.

\subsection{More about the chiral nuclear potentials}
\label{Sec3B}

The most recent generation of $NN$ potentials is derived within the formalism of ChPT. In this framework, the $NN$ interaction is governed by the (approximate) chiral symmetry
of low-energy QCD that constrains the building blocks of the $NN$ Lagrangian. 
ChPT provides a description of nuclear systems in terms of single and multiple pion exchanges (long- and medium-range components) and contact interactions between the nucleons in order to
parametrize the short-range behaviour~\cite{ Epelbaum:2008ga, Machleidt:2011zz}.
A power counting scheme, based on an expansion parameter determined by the ratio of a soft scale (usually the momentum $p$ or the pion mass $m_{\pi}$) over a hard scale (i.e. QCD energy gap $\Lambda_{\chi}$), is at the basis of the perturbative expansion ~\cite{Weinberg:1990rz, BERNARD_1995, Scherer:2002tk}. 
The free parameters of the theory are determined by reproducing data in the two-nucleon sector.

In our previous works ~\cite{Vorabbi:2015nra,Vorabbi:2017rvk, Vorabbi_2018, Vorabbi_2020, Gennari:2017yez} we applied chiral $NN$ potentials at
N$^{3}$LO (next-to-next-to-next-to-leading order) and N$^{4}$LO (next-to-next-to-next-to-next-to-leading order) to the description of proton-nucleus (but also antiproton-nucleus)
elastic scattering observables.
Despite an overall agreement with experimental data, the description of the scattering observables can still be improved, in particular
concerning the polarization quantities, like the analyzing power.

One key feature of the application of ChPT in the nuclear sector is the natural emergence, as well as the fully consistent derivation, of multi-nucleon forces. The first introduction of $3N$
forces in terms of $\pi$-exchange dynamics dates back to the seminal paper of Fujita and Miyazawa ~\cite{10.1143/PTP.17.360}, where 
a single $\pi$ is exchanged between two of the three nucleons involved. 
In ChPT, such contribution naturally arises from the structure of the Lagrangian dictated by chiral symmetry. In fact, $3N$ forces start to appear at N$^{2}$LO, whereas
at LO and NLO only $NN$ contributions are allowed. As shown in Refs. ~\cite{PhysRevC.59.53}, the $2\pi$ exchange diagram between three nucleons must be completed by two
more contributions: a one-$\pi$-exchange plus a $NN$ contact term and a $3N$ contact term. 
For more details and an explicit derivation of the relevant formulae, we refer the reader to Refs. ~\cite{PhysRevC.66.064001, Sauer_2014, Navratil2007}.

The $NN$ tuning of the parameters partially constrains the $3N$ forces.
The $2\pi$-exchange part depends on the Low Energy Constants (LEC) $c_1, c_3, c_4$ (which already appear in the $NN$ sector in the subleading $2\pi$-exchange contribution at N$^{2}$LO),
 while the other contributions depend on new LECs $c_D, c_E$ that must be fixed by three-body properties. The calibration of  $c_D, c_E$ can be 
obtained by different methods ~\cite{Machleidt_2016}:
reproducing the binding energies of $^{3}$H and $^{4}$He ~\cite{PhysRevC.73.064002}, or the neutron-doublet scattering length ~\cite{PhysRevC.66.064001}, fitting some properties of light nuclei ~\cite{PhysRevLett.99.042501}, or determining the Gamow-Teller matrix element of tritium 
$\beta$-decay ~\cite{PhysRevLett.103.102502,*PhysRevLett.122.029901}.
For an exhaustive analysis about the determination of $c_D$ and $c_E$ we 
refer the reader to Ref. ~\cite{PhysRevC.99.024313}.

In the description of scattering observables, since $3N$ forces will be approximated by Eq.~(\ref{v_3Nf}), we need a theoretical prescription to average $3N$ forces over the Fermi
sphere to produce $\braket{w_{0i}}$. In Ref. ~\cite{Holt_2010} the authors proposed a method to construct a density-dependent $NN$ force generated by $3N$ forces. 
In the present work we strictly follow this procedure and refer the reader to 
the relevant bibliography for more details. 
Such approaches, where the complexity of the $3N$ force is reduced to a density dependent $NN$ force, 
have been successfully tested by many authors, see Refs. ~\cite{Bogner:2009bt,Hammer:2012id,Hagen:2013nca} and references therein and, in particular, Ref. ~\cite{Holt_2013}, where
an optical potential for infinite systems has been derived. 
Since in finite nuclei the baryon density $\rho$ is a function of the radial coordinate, it would be necessary to find a 
prescription to fix $\rho$. We choose a different approach because the goal of our work is to investigate $3N$ forces in a broad sense. We allow $\rho$ to vary between reasonable
values going from surface-like densities to bulk-like densities. As a consequence, our theoretical predictions will be drawn as``bands" and not single lines. These bands should not be
confused with similar bands (of uncertainty) that we presented in our previous work ~\cite{Vorabbi:2017rvk} related to the errors associated with the chiral perturbative expansion.

\subsection{Practical details}
\label{sect_op}

From a practical point of view, the OP is computed in momentum space as follows ~\cite{PhysRevC.40.881,PhysRevC.41.814, PhysRevC.56.2080, PhysRevC.57.1378}
\begin{widetext}
\begin{equation}\label{fullfoldingop}
\begin{split}
U ({\bm q},{\bm K}; E ) &= \sum_{N=p,n} \int d {\bm P} \; \eta ({\bm q},{\bm K},{\bm P}) \;
t_{NN} \left[ {\bm q} , \frac{1}{2} \left( \frac{A+1}{A} {\bm K} + \sqrt{\frac{A-1}{A}} {\bm P} \right) ; E \right] \\
&\times \rho_N \left( {\bm P} + \sqrt{\frac{A-1}{A}} \frac{{\bm q}}{2} , {\bm P} - \sqrt{\frac{A-1}{A}} \frac{{\bm q}}{2} \right) \, ,
\end{split}
\end{equation}
\end{widetext}
where ${\bm q}$ and ${\bm K}$ represent the momentum transfer and the average momentum, respectively. Here ${\bm P}$ is an integration variable, $t_{NN}$ is the $NN$
$t$-matrix [Eq.~(\ref{final_t})] and $\rho_N$ is the one-body nuclear density matrix. 
The parameter $\eta$ is the M\"{o}ller factor, that imposes the Lorentz invariance of the flux when we pass from
the $NA$ to the $NN$ frame in which the $t$ matrices are evaluated. 
Finally, $E$ is the energy at which the $t$ matrices are evaluated and it is fixed at one half 
the kinetic energy of the incident nucleon in the laboratory frame. 

The calculation of the
density matrix is performed using the same approach followed in Ref.~\cite{Gennari:2017yez}, where one-body translationally invariant densities were computed within the
{\it ab initio} NCSM~\cite{BARRETT2013131} approach.
The NCSM method is based on the expansion of the nuclear wave functions in a harmonic oscillator basis and it is thus characterized by the harmonic oscillator
frequency $\hbar \Omega$ and the parameter $N_{max}$, which specifies the number of  nucleon excitations above the lowest energy configuration allowed by the Pauli principle.
In this work, the densities have been computed using $\hbar \Omega = 16$ MeV and $N_{max} = 8$ for $^{12}$C and $^{16}$O (within the importance truncated
method~\cite{PhysRevLett.99.092501,PhysRevC.79.064324}). The center-of-mass contributions have been consistently removed~\cite{Gennari:2017yez}.

For the present work we used the $NN$ chiral interactions developed by Entem {\it et al.}~\cite{Entem:2014msa,Entem:2017gor} up to the fifth order (N$^4$LO) with
a cutoff $\Lambda = 500$ MeV for both the target description and the interaction potential $V$ [cf.\ Eq.~(\ref{V_with_3B})] between the projectile and the target nucleon.
In addition to the $NN$ interaction, we also employed genuine $3N$ forces to compute the one-body densities of the target nuclei. We adopted the $3N$ chiral
interaction derived up to third order (N\textsuperscript{2}LO), which employs a simultaneous local and nonlocal regularization with the cutoff values of
$650$ MeV and $500$ MeV, respectively~\cite{Navratil2007,Gysbers2019}. For the present work we use the values $c_D = -1.8$ and $c_E = -0.31$ with the $NN$ interaction at
N$^{4}$LO \cite{Vorabbi_2020,Gysbers2019}, while with the $NN$ interaction at N$^3$LO and N$^2$LO we used the values provided in
Table I of Ref.~\cite{PhysRevC.102.024616}.
For the NCSM, the interaction is also renormalized using the similarity renormalization group (SRG) technique, which
evolves the bare interaction at the desired resolution scale $\lambda_{\mathrm{SRG}} = 2.0$ $\mathrm{fm}^{-1}$ to ensure a faster convergence of our calculations. To 
be consistent, for the evaluation of Eq.~(\ref{v_3Nf}), we employed the same values for $c_D$ and $c_E$. Finally, in the evaluation of the pure $3N$ force and of
Eq.~(\ref{v_3Nf}) we used the $c_1$, $c_3$, and $c_4$ values recommended in Ref.~\cite{Entem:2017gor}.


\section{Theoretical Predictions}
\label{sec_theoretical_predictions}

In this section we present and analyze our theoretical predictions for the elastic $NA$ scattering observables calculated with the model proposed in Sec.~\ref{sec_optical_model}.
The main goal is to evaluate the impact of genuine $3N$ forces in the description of empirical data. We refer the reader to Refs. ~\cite{Vorabbi:2015nra,Vorabbi:2017rvk, Vorabbi_2018}
for extensive analyses about the dependence on the details of $NN$ chiral potentials, convergence patterns, and error estimates at a given order of the chiral expansion.

All the theoretical results were obtained using Eq.(\ref{fullfoldingop}), where the $t_{pN}$ matrix is computed with the $pN$ chiral interaction of Ref.~\cite{Entem:2017gor}
supplemented by a density dependent $NN$ interaction  and the one-body nonlocal density matrices computed with the NCSM
method using $NN$~\cite{Entem:2017gor} and $3N$~\cite{Navratil2007,Gysbers2019} chiral interactions.

\begin{figure}[h]
\begin{center}
\includegraphics[scale=0.65]{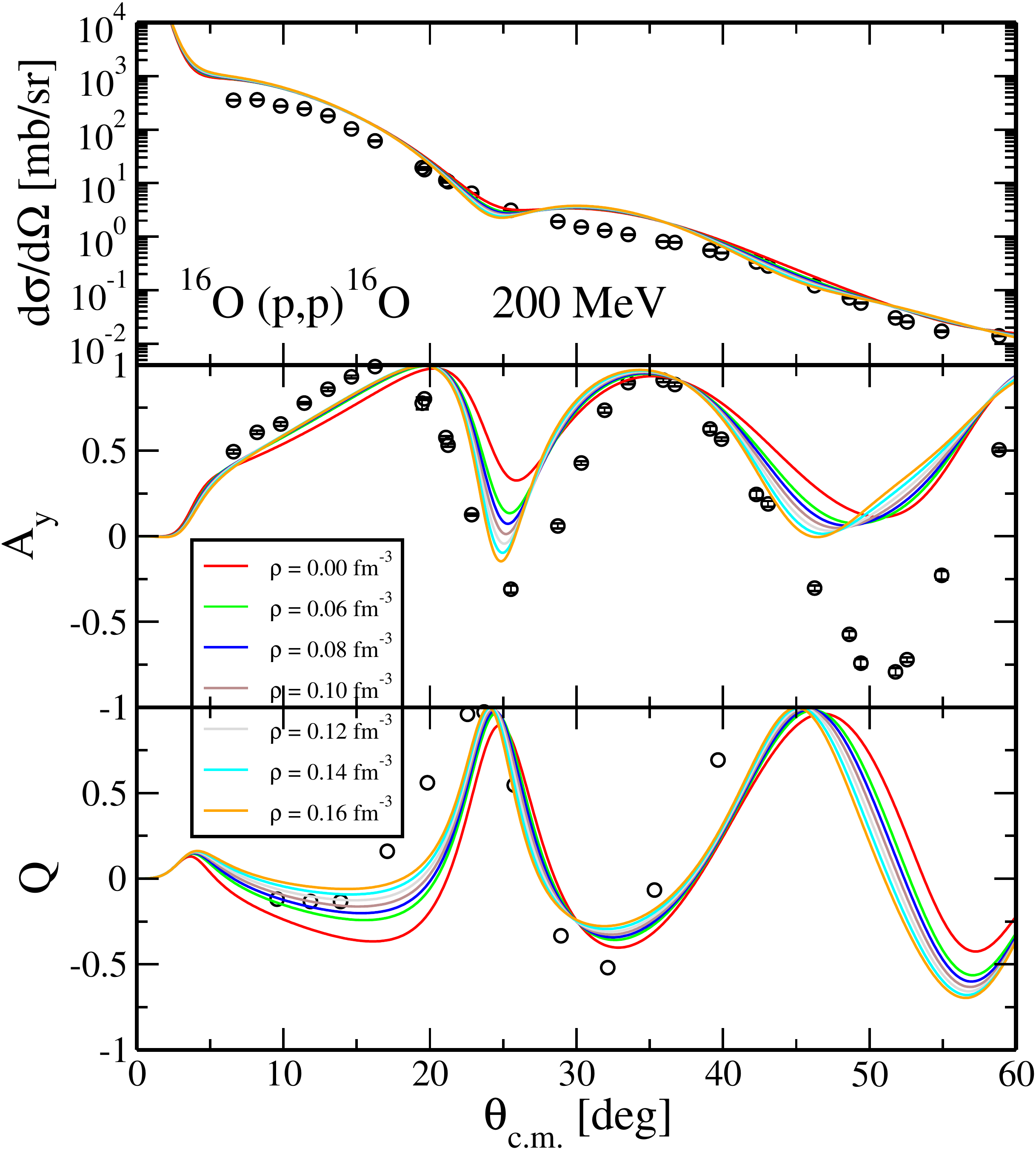}
\caption{ (Color online) 
Differential cross section $d\sigma/ d\Omega$, analyzing power $A_y$, and spin rotation $Q$ as functions of the center-of-mass scattering angle $\theta_{c.m.}$  for elastic
proton scattering off  $^{16}$O at a laboratory energy of 200 MeV. 
The results were obtained using Eq.(\ref{fullfoldingop}), where the $t_{pN}$ matrix is computed with the $pN$
chiral interaction of Ref.~\cite{Entem:2017gor} supplemented by a density dependent $NN$ interaction (with $ 0.0$ fm$^{-3} \le \rho \le 0.16$ fm$^{-3}$) and the one-body nonlocal
density matrices computed with the NCSM method using $NN$~\cite{Entem:2017gor} and $3N$~\cite{Navratil2007,Gysbers2019} chiral interactions.
Experimental data from Refs.~\cite{PhysRevC.47.1615,PhysRevC.31.1}.\label{fig_16O_200MeV}
}
\end{center}
\end{figure}

In Fig. \ref{fig_16O_200MeV} we display the calculated differential cross section $d\sigma/ d\Omega$, analyzing power $A_y$, and spin rotation $Q$ as functions of the center-of-mass
scattering angle $\theta_{c.m.}$ for elastic proton scattering off $^{16}$O at a laboratory energy of 200 MeV in comparison with the empirical data ~\cite{PhysRevC.47.1615,PhysRevC.31.1}.
The set of curves show the impact of genuine $3N$ forces 
with increasing values of the matter density $\rho$ (with $ 0.0$ fm$^{-3} \le \rho \le 1.6$ fm$^{-3}$) starting from the case with only $NN$ contributions ($\rho=0$). The effects of genuine $3N$ forces turn out to be negligible for the differential cross section, where all curves are basically on top of each other, and are larger  for polarization observables, where the $3N$ contributions $\braket{w_{0i}}$ improves the agreement with the experimental data, in particular, there is a strong improvement in the description of first minimum of $A_y$. 

\begin{figure}[h]
\begin{center}
\includegraphics[scale=0.65]{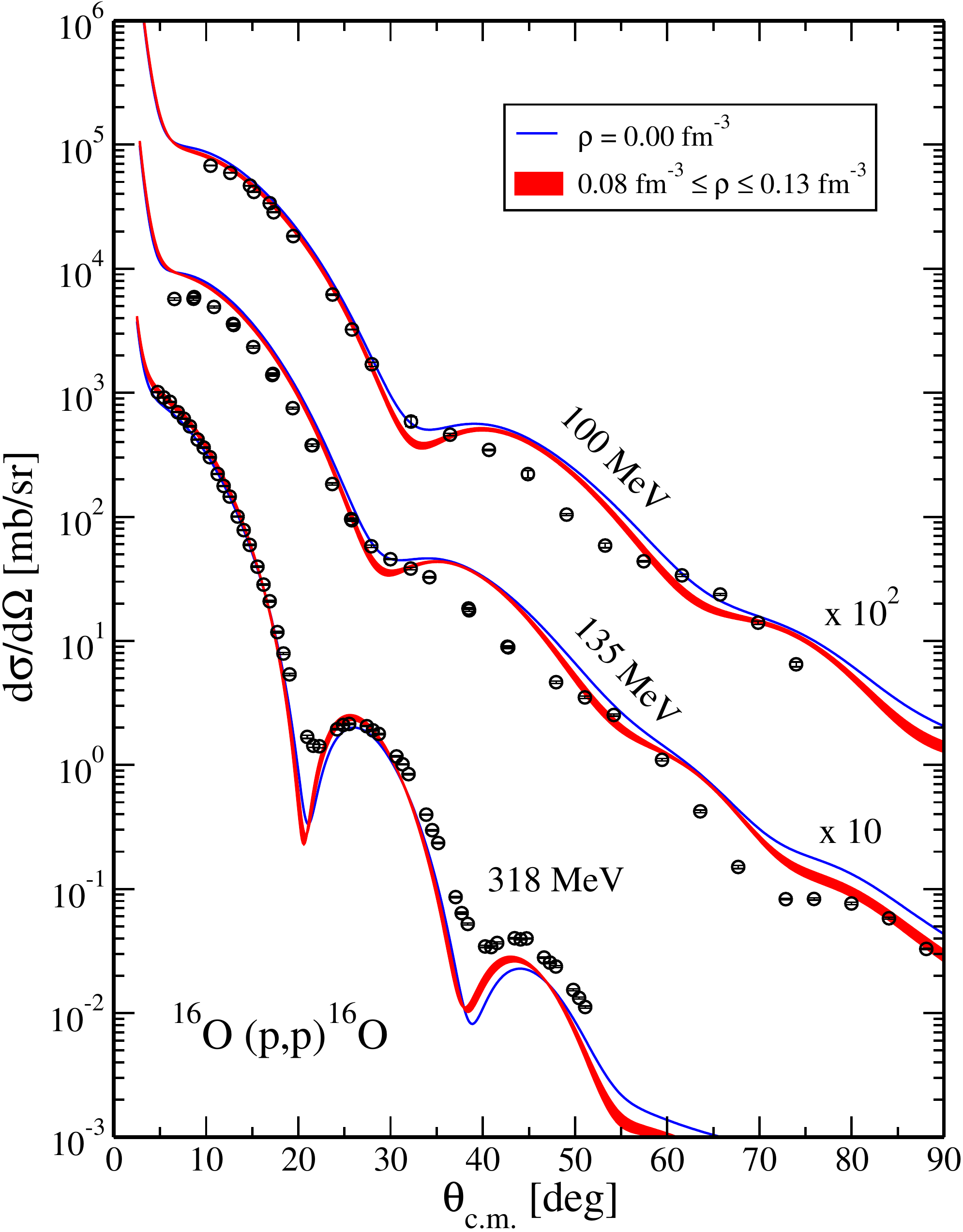}
\caption{ (Color online) Differential cross sections $d\sigma/ d\Omega$ as a function of the center-of-mass scattering angle for elastic proton scattering off $^{16}$O at different
energies (100, 135 and 318 MeV). 
The bands show the results obtained using Eq.(\ref{fullfoldingop}), where the $t_{pN}$ matrix is computed with the $pN$ chiral interaction of
Ref.~\cite{Entem:2017gor} supplemented by a density dependent $NN$ interaction 
(with $ 0.08$ fm$^{-3} \le \rho \le 0.13$ fm$^{-3}$) 
and the one-body nonlocal density matrices computed with the NCSM method using $NN$~\cite{Entem:2017gor} and $3N$~\cite{Navratil2007,Gysbers2019} chiral interactions.
The solid (blue) lines are obtained with $\rho= 0$ fm$^{-3}$. 
Experimental data from Refs.~\cite{PhysRevC.47.1615,PhysRevC.31.1, PhysRevC.39.1222, PhysRevC.43.1272}.
\label{fig_16O_sigma} }
\end{center}
\end{figure}

In Fig.~\ref{fig_16O_sigma} we show the differential cross sections as functions of the center-of-mass scattering angle for elastic proton scattering off $^{16}$O at different energies
(100, 135 and 318 MeV). Our theoretical predictions are depicted as bands, as explained in Sec.~\ref{Sec3B}, in order to show how $3N$ contributions affect the observable varying
the matter density $\rho$ within reasonable estimates ($ 0.08$ fm$^{-3} \le \rho \le 0.13$ fm$^{-3}$). For each energy, the addition of $\braket{w_{0i}}$ does not appreciably change
the behavior and the magnitude of $d\sigma/ d\Omega$ as a function of the scattering. The agreement with empirical data is good, in particular for $\theta \le 50^o$, where our
calculations nicely reproduce the minima of the cross sections.

\begin{figure}[h]
\begin{center}
\includegraphics[scale=0.65]{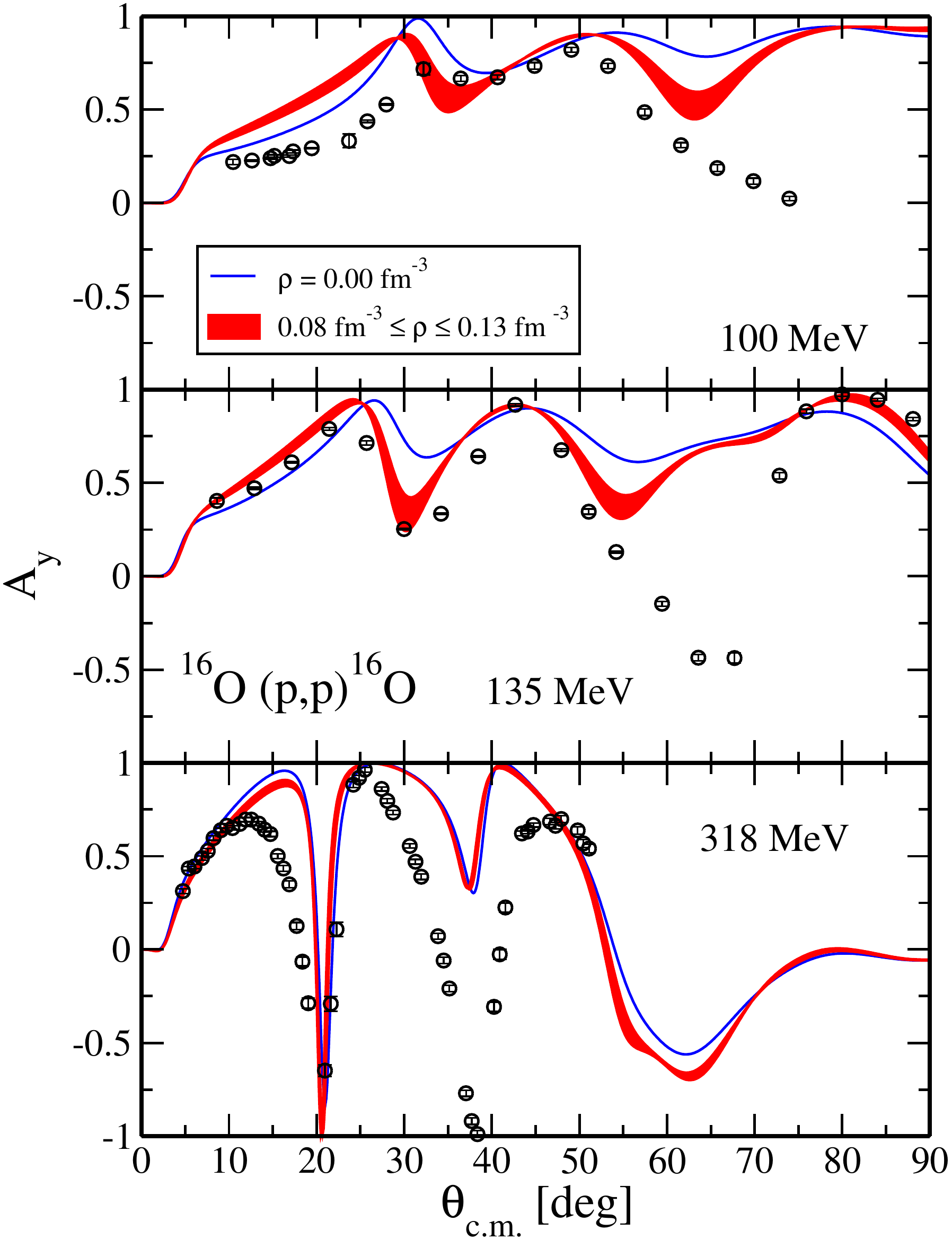}
\caption{ (Color online) The same as in Fig.~\ref{fig_16O_sigma} but for the analyzing power $A_y$.
Experimental data from Refs.~\cite{PhysRevC.47.1615,PhysRevC.31.1, PhysRevC.39.1222, PhysRevC.43.1272}.
\label{fig_16O_Ay} }
\end{center}
\end{figure}

In Fig.~\ref{fig_16O_Ay} we plot the analyzing power $A_y$ as a function of the center-of-mass scattering angle for the same nucleus at the same energies
(100, 135 and 318 MeV). As in the previous figure, our theoretical predictions are shown as  bands. The comparison with the calculations with only $NN$ interactions (solid curves) show that the effects of genuine $3N$ forces are larger for polarization observables. We do not show results for the spin rotation $Q$ because there are no empirical data at these energies. For data at low energy,
the contribution of genuine $3N$ forces generally improves the description of empirical data, in particular of the shape of $A_y$, while minima positions are less affected. This is evident from the results at 135 MeV. The results at 318 MeV are less sensitive to the contribution  of genuine $3N$ forces. 
The results at 100 MeV deserve a special comment, since at this energy $A_y$ computed with $\rho = 0$ fm$^{-3}$ seems to provide a better description of the data for $\theta_{c.m.} \lesssim 30^{\circ}$. This different behavior, compared to the other cases in the figure, can be ascribed to the impulse approximation used to derive Eq.~(\ref{fullfoldingop}). At 100 MeV medium effects can be important and the validity of the impulse approximation can be put into question. The experimental differential cross section at 100 MeV in Fig.~\ref{fig_16O_sigma} is anyhow reasonably described by the model. We note that even in the case of the cross section at 100 MeV the impact of the $3N$ contribution, although  small, does not improve but rather worsens the agreement with the experimental data. 

\begin{figure}[h]
\begin{center}
\includegraphics[scale=0.65]{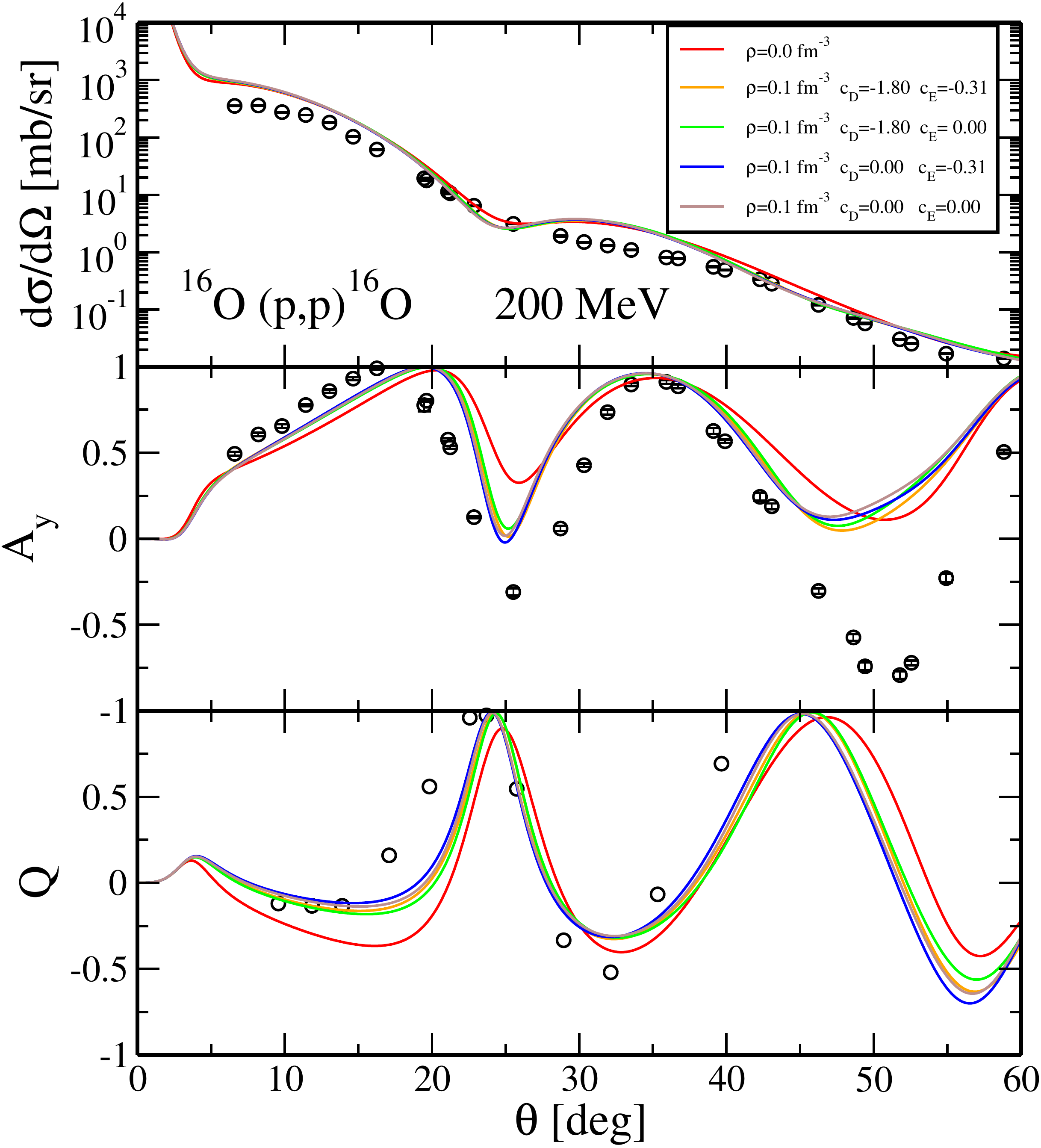}
\caption{(Color online) Differential cross section $d\sigma/ d\Omega$, analyzing power $A_y$, and spin rotation $Q$ as functions of the center-of-mass scattering angle for elastic
proton scattering off $^{16}$O at a laboratory energy of 200 MeV for different combinations of the low-energy constants $c_D$ and $c_E$. The red curve is the theoretical prediction
with only the $pN$ chiral interaction of Ref.~\cite{Entem:2017gor}, while the other curves are generated switching on and off the effective $3N$ contributions. 
Experimental data from Refs.~\cite{PhysRevC.47.1615,PhysRevC.31.1}.
\label{fig_16O_200MeV_cD_cE} }
\end{center}
\end{figure}

We conclude the analysis of the results for  $^{16}$O by showing in Fig.~\ref{fig_16O_200MeV_cD_cE} a comparison of the differential cross section, analyzing power, and spin rotation as functions of
the center-of-mass scattering angle for different combinations of the low-energy constants $c_D$ and $c_E$. The theoretical prediction with only the $pN$ chiral
interaction of Ref.~\cite{Entem:2017gor} (red lines) are compared in the figure with the results generated switching on and off the effective $3N$ contributions. Since the dependence on $c_D$ and $c_E$ is
very weak, it is reasonable to state that the main contribution of the $3N$ force comes from the $3N$-2$\pi$ exchange diagrams, that depend only on $c_1, c_3$, and $c_4$.

\begin{figure}[h]
\begin{center}
\includegraphics[scale=0.65]{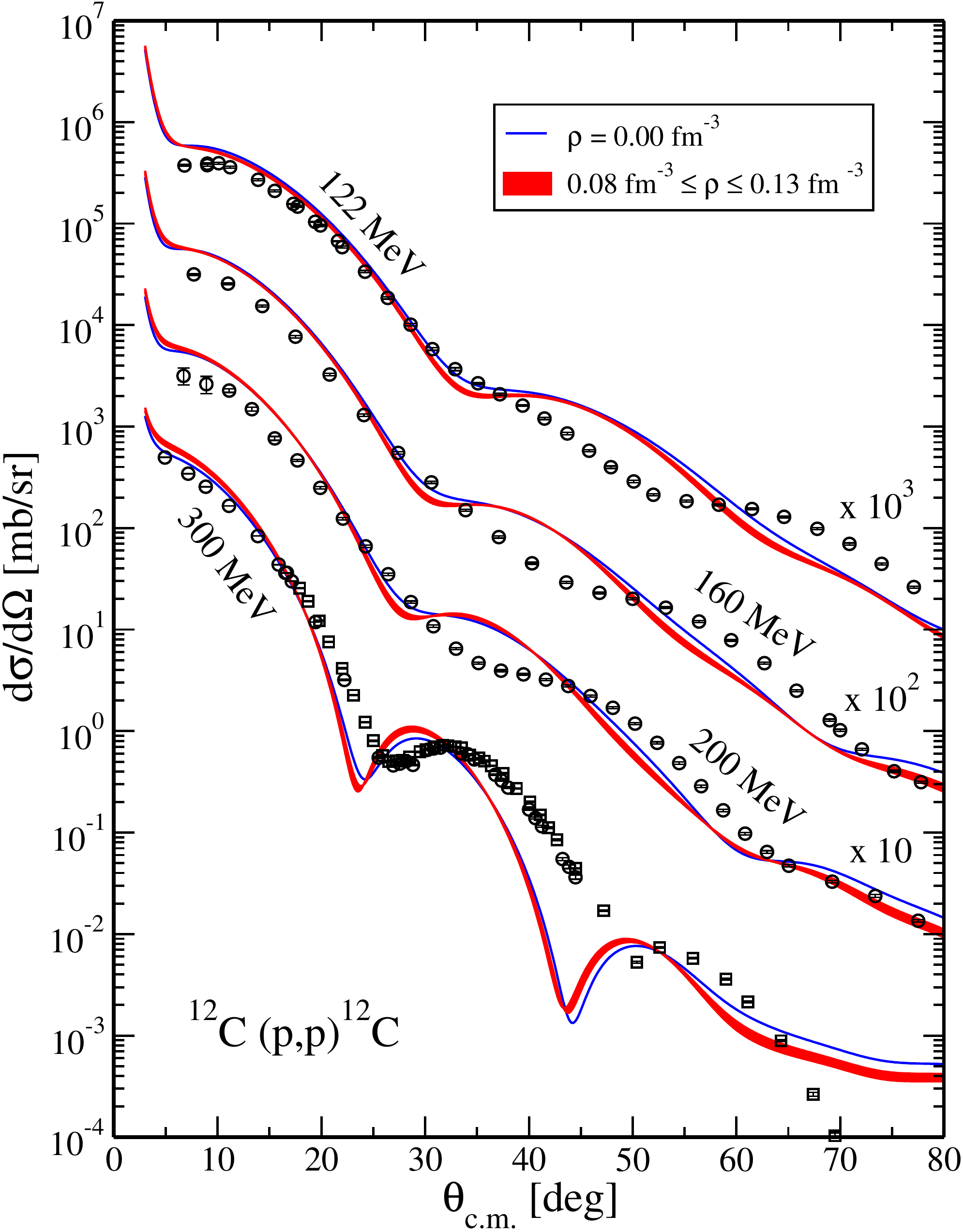}
\caption{ (Color online) The same as in Fig. \ref{fig_16O_sigma} but for $^{12}$C and for different energies (122, 160, 200, and 300 MeV).
Experimental data from Refs.~\cite{PhysRevC.21.2147, PhysRevC.27.459, PhysRevC.23.616, PhysRevC.81.054604,PhysRevC.31.1569}.
\label{fig_12C_sigma} }
\end{center}
\end{figure}

\begin{figure}[h]
\begin{center}
\includegraphics[scale=0.65]{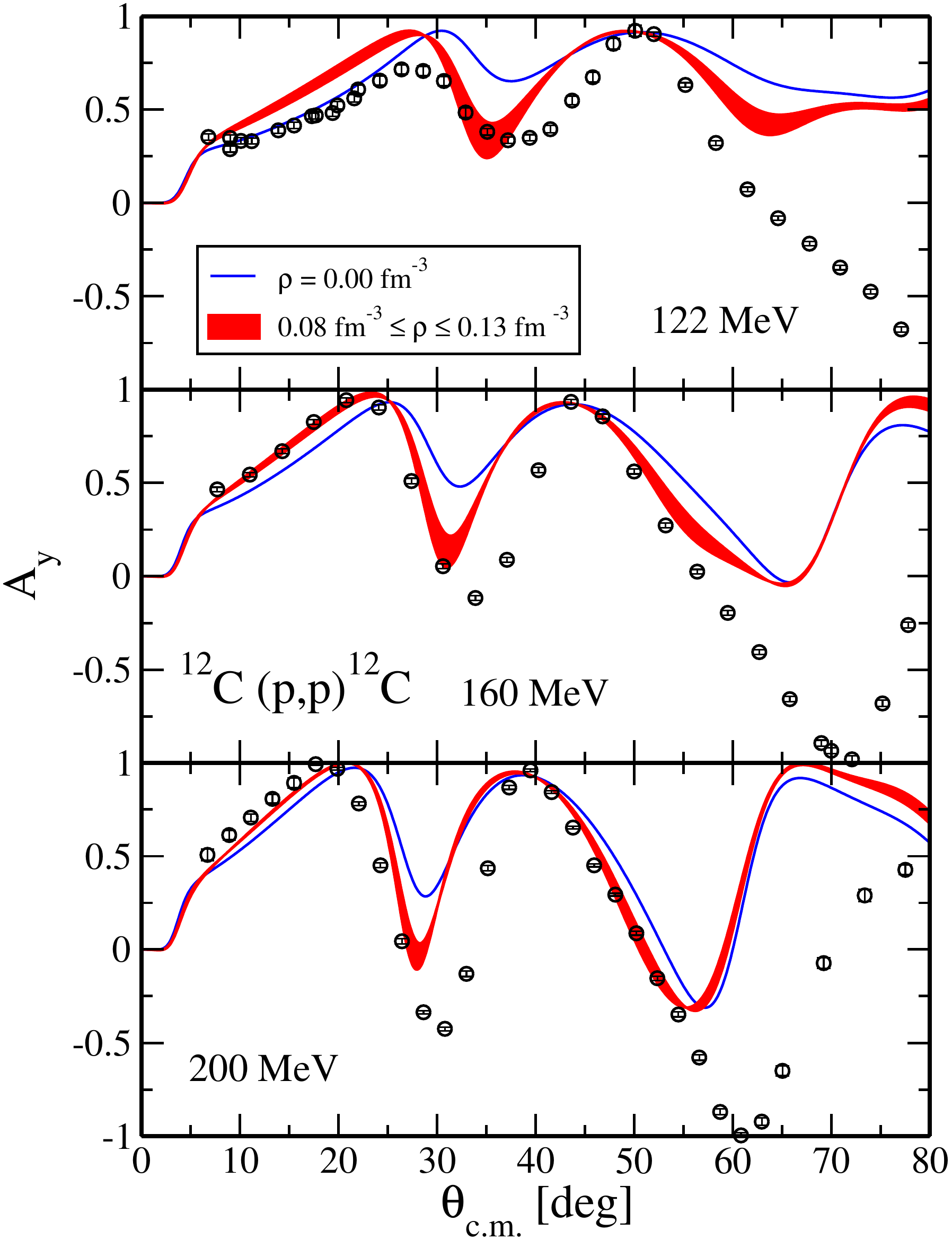}
\caption{ (Color online) The same as in Fig. \ref{fig_16O_Ay} but for $^{12}$C and for different energies (122, 160, and 200 MeV).
Experimental data from Refs.~\cite{PhysRevC.21.2147, PhysRevC.27.459, PhysRevC.23.616}.
\label{fig_12C_Ay} }
\end{center}
\end{figure}

We continue our analysis with the results for $^{12}$C: we plot the differential cross section (Fig. \ref{fig_12C_sigma}) and analyzing power (Fig. \ref{fig_12C_Ay}) as functions of the center-of-mass
scattering angle at different energies (122, 166, 200, and 300 MeV) in comparison with the experimental
data ~\cite{PhysRevC.21.2147, PhysRevC.27.459, PhysRevC.23.616, PhysRevC.81.054604,PhysRevC.31.1569}. No results are shown for the spin rotation because no experimental data
at these energies are available. In the carbon case we observe the same pattern and as for oxygen and we can draw the same conclusions.
Genuine $3N$ forces appear to have a very small impact on the cross sections,  for all the considered energies of the projectile, and clearly improve the description of the experimental data for polarization observables. The first minimum of $A_y$ is satisfactorily reproduced both in respect to the angular dependence and the magnitude.

\begin{figure}[h]
\begin{center}
\includegraphics[scale=0.65]{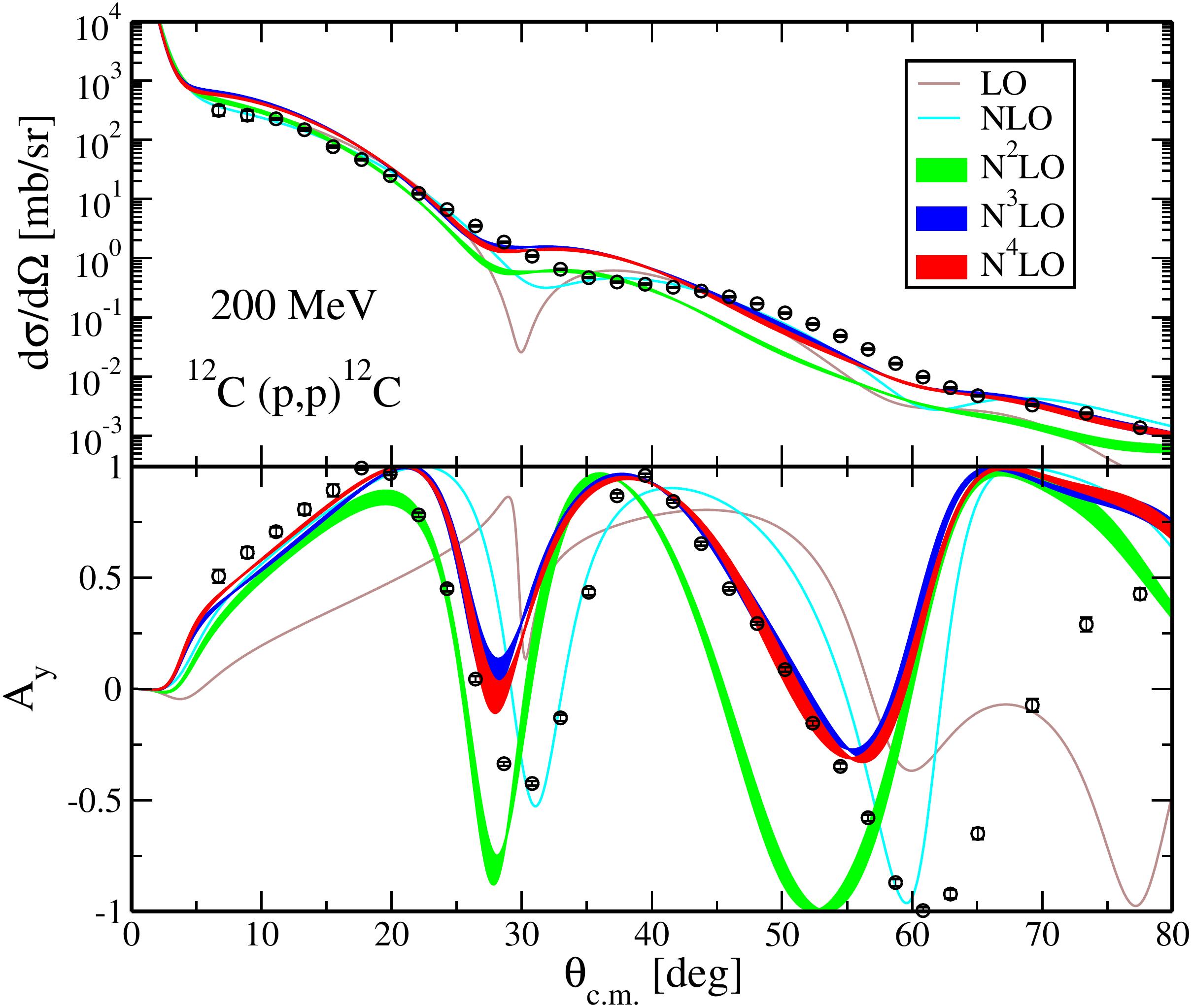}
\caption{ (Color online) Differential cross sections $d\sigma/ d\Omega$ as a function of the center-of-mass scattering angle for elastic proton scattering off $^{12}$C at 200 MeV
at different orders of the chiral expansion: LO (brown curve), NLO (cyan curve), N$^{2}$LO (green band), N$^{3}$LO (blue band), and N$^{4}$LO (red band). Since $3N$ forces
start to appear at N$^{2}$LO, at lower orders they are not included. The bands are obtained when the matter density at which the $3N$ contributions are calculated is allowed to
vary in the interval $0.08$ fm$^{-3} \le \rho \le 0.13$ fm$^{-3}$.
Experimental data from Refs.~\cite{PhysRevC.21.2147, PhysRevC.27.459, PhysRevC.23.616, PhysRevC.81.054604,PhysRevC.31.1569}.
\label{fig_12C_200MeV_orders} }
\end{center}
\end{figure}

For the carbon case we also performed an order-by-order analysis in terms of the chiral order expansion. In Fig. \ref{fig_12C_200MeV_orders} we show the differential cross sections
$d\sigma/ d\Omega$ as functions of the center-of-mass scattering angle for elastic proton scattering off $^{12}$C at 200 MeV at different orders of the chiral expansion.
Since $3N$ forces start to appear at N$^{2}$LO, at lower orders they are not included and the predictions are plotted as lines and not bands. Starting from N$^{2}$LO, the bands are
obtained when the matter density at which the $3N$ contributions are calculated is allowed to vary in the interval $ 0.08$ fm$^{-3} \le \rho \le 0.13$ fm$^{-3}$. At each order, 
we refitted $c_D$ and $c_E$ to ensure consistency ~\cite{PhysRevC.102.024616}, following the same prescriptions explained in the previous section. To ensure complete consistency,
we used the same potentials both in the NCSM calculations and in the projectile-target interaction. As also shown in our previous
papers ~\cite{Vorabbi:2015nra,Vorabbi:2017rvk, Vorabbi_2018}, already at order N$^{3}$LO a good degree of convergence is achieved.

\begin{figure}[h]
\begin{center}
\includegraphics[scale=0.65]{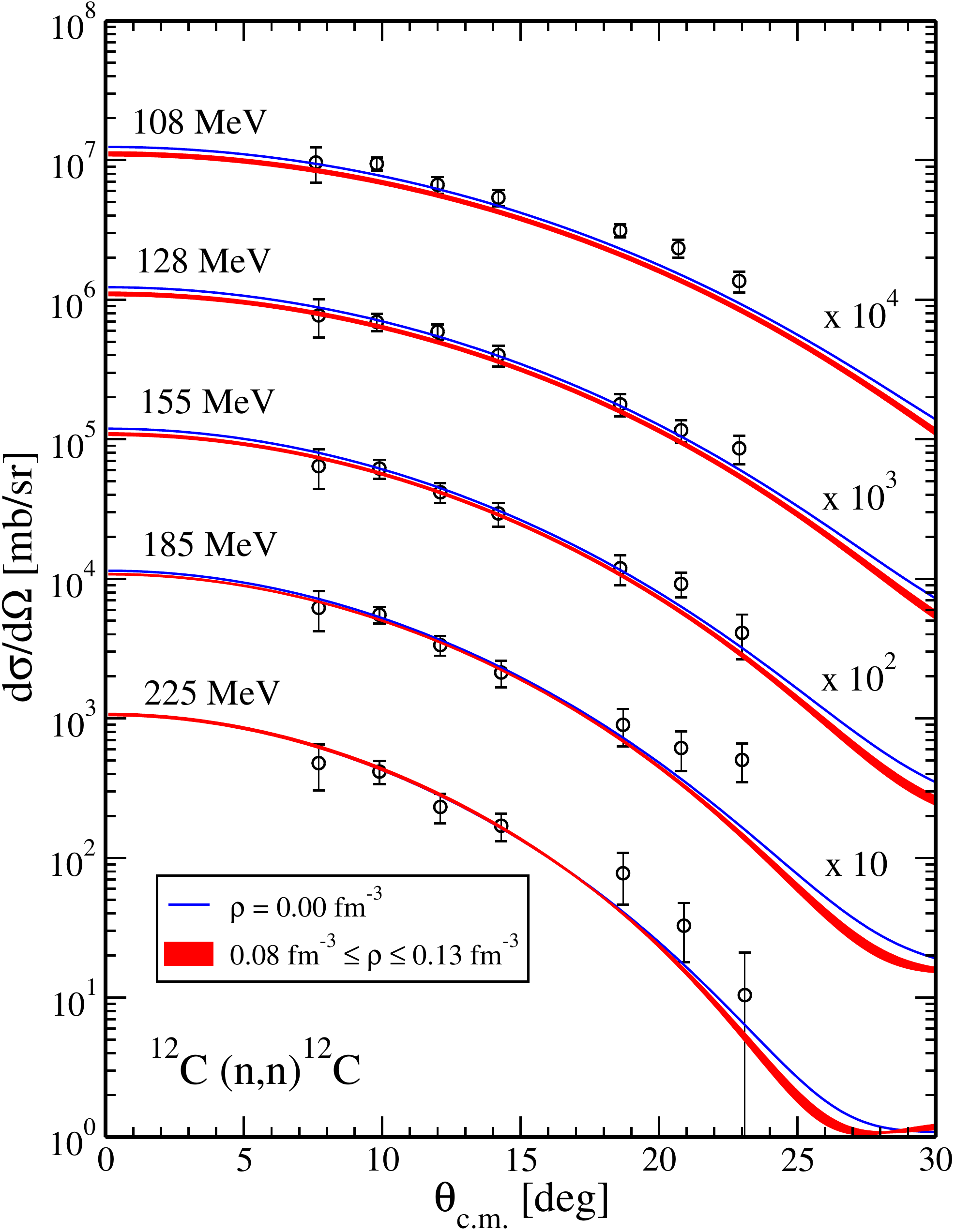}
\caption{ (Color online) 
\label{fig_n_12C} Differential cross sections $d\sigma/ d\Omega$ as a function of the center-of-mass scattering angle for elastic neutron scattering off $^{12}$C at different energies
(108, 128, 155, 185, and 225 MeV). The results were obtained using Eq.(\ref{fullfoldingop}), where the $t_{nN}$ matrix is computed with the $nN$ chiral interaction of
Ref.~\cite{Entem:2017gor} supplemented by a density dependent $NN$ interaction (with $ 0.08$ fm$^{-3} \le \rho \le 0.13$ fm$^{-3}$) and the one-body nonlocal density matrices
computed with the NCSM method using $NN$~\cite{Entem:2017gor} and $3N$~\cite{Navratil2007,Gysbers2019} chiral interactions.
Experimental data from Refs.~\cite{PhysRevC.70.054613}.}
\end{center}
\end{figure}

Finally, we also checked our approach for neutron elastic scattering off $^{12}$C.  In Fig. \ref{fig_n_12C} we show the
differential cross sections $d\sigma/ d\Omega$ as a function of the center-of-mass scattering angle at different energies (108, 128, 155, 185, and 225 MeV) in comparison with the 
experimental data~\cite{PhysRevC.70.054613}. The agreement with the empirical data is overall good, for all the energies considered. The inclusion of $3N$ forces  does not appreciably change the results obtained  with only the $NN$ chiral potential, reinforcing our previous conclusions, drawn from the results for elastic proton scattering, that genuine $3N$ forces give only a small contribution to the differential cross section. They  seem to provide sizable contributions only for observables related to polarized particles. No empirical polarization data are available for neutron elastic scattering off $^{12}$C and we do not show results for polarization observables.

\section{Conclusions}
\label{sec_conclusions}

In a previous papers we obtained an intermediate energy microscopic OP for elastic nucleon-nucleus scattering from chiral potentials. The OP was derived at the first order term
of the Watson multiple scattering theory and adopting the impulse approximation. 
The final expression of the OP~\cite{Gennari:2017yez} was a folding integral between the $NN$ $t$ matrix and the one-body density of the target.
We used the $3N$ force only in the calculation of the target density while the $t$ matrix, that represents the dynamic part of the OP, was computed
with only the $NN$ interaction. Of course, for a more consistent calculation, the $3N$ force should be included in the dynamic part of the OP as well. Unfortunately, the exact
treatment of the $3N$ force involves multiple scattering terms of the projectile with the target nucleons that would make the calculation too difficult for our current capabilities and that have been neglected. 

The goal of the present work is to introduce a suitable approximation that allows us to
include the $3N$ interaction also in the dynamic part of the OP already at the level of single-scattering approximation between the projectile and the target nucleon. Our technique is based on
averaging the $3N$ force over the Fermi sphere and thus defining a density dependent $NN$ interaction which acts as a medium correction for the bare $NN$ potential.
This treatment naturally extends the previous expression of the OP and allows a direct comparison of our new and old results.

We considered $^{12}$C and $^{16}$O as case studies and we computed the differential cross section and the polarization observables for different energies
of the incoming protons and neutrons. Our finding is that the contribution of the $3N$ interaction in the dynamic part of the OP is very small and almost negligible on the
differential cross section, while it is sizable on the polarization observables where it improves the agreement with the experimental data. 

Moreover, switching on and off the values of the
$c_D$ and $c_E$ constants in the $3N$ interaction allowed us to identify the diagram that mostly contributes to the final $3N$ force, {\it i.e.}, the 2$\pi$ exchange term.

Finally, we checked the order by order convergence of the chiral expansion comparing results at different orders, refitting,  at each order, the values of the $c_D$ and $c_E$ constants and  found that, especially at LO and NLO, the results are pretty erratic and they start to reach convergence only at  N$^3$LO, in agreement with our previous calculations.




\section{Acknowledgements}

The work at Brookhaven National Laboratory was sponsored by the Office of Nuclear Physics, Office of Science of the U.S. Department of Energy under
Contract No. DE-AC02-98CH10886 with Brookhaven Science Associates, LLC.
The work at TRIUMF was supported by the NSERC Grant No. SAPIN-2016-00033. TRIUMF receives federal funding via a contribution agreement with the National
Research Council of Canada.
Computing support came from an INCITE Award on the Summit supercomputer of the Oak Ridge Leadership Computing Facility (OLCF) at ORNL, from Westgrid and
Compute Canada.
The work by R.M.\ was supported in part by the U.S. Department of Energy
under Grant No.~DE-FG02-03ER41270.


%

\end{document}